\documentstyle[12pt]{article}
\newcommand{\be}{\begin{equation}}
\newcommand{\ee}{\end{equation}}
\newcommand{\bea}{\begin{eqnarray}}
\newcommand{\ena}{\end{eqnarray}}
\newcommand{\beano}{\begin{eqnarray*}}
\newcommand{\enano}{\end{eqnarray*}}
\newcommand{\sect}[1]{\setcounter{equation}{0}\section{#1}}
\newcommand{\vs}[1]{\rule[- #1 mm]{0mm}{#1 mm}}

\newcommand{{\cg}}{\mbox{$\cal{G}$}}

\newcommand{\R}{\mbox{\hspace{.04mm}\rule{0.2mm}{2.8mm}\hspace{-1.5mm} R}}
\newcommand{\C}{\mbox{\hspace{1.24mm}\rule{0.2mm}{2.5mm}\hspace{-2.7mm} C}}

\newcommand{\Z}{\mbox{$Z\hspace{-2mm}Z$}} 

\begin{document}
\renewcommand{\thefootnote}{\fnsymbol{footnote}} \newpage
\pagestyle{empty}
\setcounter{page}{0}
\rightline{September 94}

\vs{20}

\centerline{\LARGE{\bf Triangular dissections, aperiodic tilings}}
\vs{2} \centerline{\LARGE{\bf and}}\vs{2}
\centerline{\LARGE{\bf Jones algebras}}

\vs{10}

\centerline{{\Large R. Coquereaux}}

{\it \centerline{Centre de Physique Th\'eorique}
 \centerline{CNRS Luminy - Case 907}
\centerline{F 13288 Marseille Cedex 9 (France)}}

\vs{10}

{\bf Abstract :}
The Brattelli diagram associated with a given bicolored
Dynkin-Coxeter graph of type $A_n$ determines planar fractal sets
obtained by infinite dissections of a given triangle. All triangles
appearing in the dissection process have angles that are multiples of
$\pi/ (n+1).$ There are usually several possible infinite dissections
compatible with a given $n$ but a given one makes use of $n/2$
triangle types if $n$ is even. Jones algebra with index $[ 4 \
\cos^2{\pi \over n+1}]^{-1}$ (values of the discrete range) act
naturally on vector spaces associated with those fractal sets.
Triangles of a given type are always congruent at each step of the
dissection process. In the particular case $n=4$, there are isometric
and the whole structure lead, after proper inflation, to aperiodic
Penrose tilings. The ``tilings'' associated with other
values of the index are discussed and shown to be encoded by
equivalence classes of infinite sequences (with appropriate
constraints) using $n/2$ digits (if $n$ is even) and generalizing the
Fibonacci numbers. \vs{7}


\bigskip

To appear in Advances in Applied Mathematics

\vfill

\vs{4}

\begin{minipage}{4cm}
September 1994\\
CPT - 94 / P.3020
\end{minipage}
\hfill


\newpage
\pagestyle{plain}

\indent

\sect{ Introduction.}\vs{10}
The aim of this paper is to show how to associate infinite
(fractal-like) dissections of triangles to Brattelli diagrams
describing Jones algebras. We limit our discussion to those algebras
related with the $A_n$ series. \par
 In section 2 , before describing the general features of the
construction, we study in details the case $n=6$. As we shall see,
there are essentially three different solutions of the dissection
problem for this value of $n$. What we mean by a ``solution'' will
become clear later. We then also discuss the relatively trivial cases
$n=3$ and $n=4$. The choice $n=4$ is actually not really trivial,
since it is related to Penrose tilings, but the determination of the
triangle types appearing in the dissection process is quite simple
(only one solution). We finally give (without pictures) the six
solutions corresponding to the case $n=8$.  \par
 In section 3, we remind the reader what Jones algebras are, and how
to relate those associated with $A_n$ Dynkin diagrams to the
fractal sets built in the previous section.\par In section 4,
following \cite{Connes1}, we first explain why the Jones algebra
associated with the Dynkin diagram of $A_4$ can be considered as a
``non-commutative'' description of the set of Penrose tilings. We
then return to the case of $A_n$ and introduce (finite or infinite) sequences
that encode paths on Brattelli diagrams. We show finally how such sequences are
in one to one correspondance with the triangles appearing in the
dissection process described in the first part.\par In section 5, we
make a few remarks relating our fractal dissection of triangles to the
projection technique used, for instance, in the theory of
quasicrystals. \bigskip

\sect{ Triangular dissections associated with $A_n$
diagrams.}\vs{10}

 Let $A_n$ be a Dynkin-Coxeter diagram, i.e. a
sequence of  $n$ vertices linked by single edges. We
shall actually consider bi-colored graphs of the
following type: odd and even vertices will be decorated
with different colors. We now fold the diagram in such a way
 that odd vertices are on the
first line and even vertices on the second line, as in
Fig.1 (example of $A_6$). We then attach the integer $1$ to the
 vertices of the first line and label the
second line with $1$'s or $2$'s by imposing downward propagation
 and conservation of the sum of integers
(2=1+1), like in the usual Pascal's triangle.\par
To the previous bicolored simply connected graph (drawn on
two lines), we now attach an infinite tower called the
associated Brattelli diagram. Its construction is very
elementary: We first reflect the previous two lines with
respect to the horizontal and propagate the whole
structure downward by the same procedure. The vertices are
then labeled with integers by using downward
propagation (as in Fig.2), like in the Pascal's
triangle.\par
It is convenient to parameterize lines of this diagram by
an integer $t$. We assume that $t=1$ corresponds to the
first line of the diagram, $t=2$ to the second line, etc.
Notice that the numbers labeling the vertices also
correspond to the numbers of paths originating from the
top line, propagating downward, and ending on the chosen
vertex. For this reason, these numbers can also be
considered as parameterizing random walks on a Dynkin
diagram and $t$ as a discrete time parameter.\par
It is clear that such a construction is not
limited to Dynkin-Coxeter diagrams of type $A_n$ but we
shall not study more sophisticated examples in the present
paper. The definition of a Brattelli diagram has its
origin in the theory of $\C^*$-algebras \cite{Bra} and we shall
actually need a few results from this theory in section 3,
but the above elementary definition of a Brattelli diagram
is sufficient for our purposes.\par
 Rather than working with
an arbitrary value of $n$ we decide to explain this construction in
the case $n=6$. Cases $n$ even and $n$ odd are actually slightly
different. We shall return
to general values of $n$ at the end of this section.\par
We take $n=6$. Setting $l=n+1=7$, we consider triangles
with angles that are multiples of $\pi/7$. There is
only a finite number of such triangles and they are
obtained by decomposing $7$ into a sum of three integers.
Here are the possibilities:
\begin{eqnarray*}
 7 &= 1 + 1 + 5 \qquad\qquad &\alpha \doteq \langle 1,1,5 \rangle  \\
 7 &= 2 + 2 + 3 \qquad\qquad &\beta \doteq \langle 2,2,3 \rangle   \\
 7 &= 3 + 3 + 1 \qquad\qquad &\gamma \doteq \langle 3,3,1 \rangle \\
 7 &= 4 + 2 + 1 \qquad\qquad&\tilde \pi \doteq \langle 4,2,1 \rangle
\\
 7 &= 4 + 1 + 2 \qquad\qquad &\pi \doteq \langle 4,1,2 \rangle
\end{eqnarray*}

The first three triangles $\alpha$, $\beta$, $\gamma$ are
isosceles and the last two $\pi$ and $\tilde \pi$ are
mirror images. We label the three angles in
trigonometrical order, for example
$\pi = \langle 1,2,4 \rangle = \langle 4,1,2 \rangle
= \langle 2,4,1 \rangle \neq \tilde \pi$. These five triangles
appear in the heptahedron (Fig.3).  \par Each of these triangles can
be decomposed into a union of two triangles belonging to the above
family. A straightforward analysis leads to the following
disintegration table
\begin{eqnarray*}
\alpha &\rightarrow \alpha + \pi,\quad \tilde\pi + \gamma ,\quad \pi +
\gamma,\quad \tilde\pi + \alpha \\ \beta  &\rightarrow \beta + \pi,\quad \pi +
\gamma ,\quad \tilde\pi + \gamma,\quad \tilde\pi + \beta \\ \gamma
&\rightarrow  \pi + \gamma ,\quad \tilde\pi + \gamma ,
\quad \beta + \alpha,\quad
\alpha + \beta\\ \pi    &\rightarrow \alpha + \tilde\pi,\quad \alpha +
\beta ,\quad \tilde\pi + \beta,\quad \tilde\pi + \gamma \\ \tilde\pi
&\rightarrow \alpha + \pi,\quad \alpha + \beta ,
\quad \pi + \beta,\quad \pi + \gamma
\end{eqnarray*} We now want to interpret edges of the Brattelli
diagrams as describing decomposition of triangles. The whole set of
equations will be determined by the first three lines of this
diagram. We first label the first two lines (i.e. the Coxeter-Dynkin
diagram) with six unknown parameters
 $x \quad y\quad z\quad t\quad u\quad v$, in this order,
 by following the diagram.
We impose the following constraint: Triangle types appearing at step
$t$ of the dissection process should be the same as those appearing
at step $t-2$. It may be that more general rules could also give rise
to interesting patterns but we shall impose this constraint in the
present paper. This implies, in particular, that the third line
has to be labeled as the first (with \, $x\quad z\quad u$), see
Fig.4. These parameters will ultimately be identified with particular
triangles. The pattern of edges impose the following constraints:
\begin{eqnarray*}
x &\rightarrow & y \\
z &\rightarrow & y + t \\
u &\rightarrow & t + v \\
y &\rightarrow & x + z \\
t &\rightarrow & z + u \\
v &\rightarrow & u \end{eqnarray*}
Therefore $x=y$ and $u = v$. We are left with four
unknown parameters $x \quad z\quad  t\quad  u$  and four
constraints
\begin{eqnarray*}
z &\rightarrow x + t \\
u &\rightarrow t + u \\
x &\rightarrow x + z \\
t &\rightarrow z + u \\
\end{eqnarray*}
In order to solve these constraints, we have to make use
of the disintegration table for triangles $\alpha$,
$\beta$, $\gamma$, $\pi$ and $\tilde\pi$ given before.
  We also take into account the fact that the set of solutions
is invariant by the transformation $x \leftrightarrow u$, $z
\leftrightarrow t$ (the corresponding dissections are exactly the same).
A rather tedious analysis (straightforward but
cumbersome) leads to the following complete set of $9$
solutions. Here, we give the list the solutions by giving
 $x\quad  z\quad  t\quad  u$, in this
order.
\begin{eqnarray*}
{x = <1,3,3>,\quad z = <1,4,2>,\quad t = <1,2,4>,\quad u = <1,3,3>} \\
{x = <1,3,3>,\quad z = <1,4,2>,\quad t = <1,2,4>,\quad u = <2,2,3>} \\
{x = <1,1,5>,\quad z = <1,2,4>,\quad t = <1,4,2>,\quad u = <1,3,3>} \\
{x = <1,1,5>,\quad z = <1,2,4>,\quad t = <1,4,2>,\quad u = <1,1,5>} \\
{x = <1,1,5>,\quad z = <1,2,4>,\quad t = <1,4,2>,\quad u = <2,2,3>} \\
{x = <1,1,5>,\quad z = <1,4,2>,\quad t = <1,2,4>,\quad u = <2,2,3>} \\
{x = <1,3,3>,\quad z = <1,2,4>,\quad t = <1,4,2>,\quad u = <1,1,5>} \\
{x = <1,3,3>,\quad z = <1,2,4>,\quad t = <1,4,2>,\quad u = <2,2,3>} \\
{x = <2,2,3>,\quad z = <1,2,4>,\quad t = <1,4,2>,\quad u = <2,2,3>}
\end{eqnarray*}

We may impose the following extra condition: The
three types of triangles that appear at a given line should
be of three different types, i.e. $x\neq z$,  $x\neq u$ and  $z\neq u$.
 This arbitrary
requirement can be relaxed it if we can use color or
gray levels in the pictures. Such a condition excludes, for
example, the first solution
$${x = <1,3,3> = \gamma,\quad z = <1,4,2> = \tilde\pi,
\quad t = <1,2,4> = \pi,\quad u = <1,3,3>=\gamma}$$
as well as the fourth and the last. Notice
that these excluded solutions minimize the set of ``proto-triangles'' (the
elementary triangle types used in the construction). With the above condition,
we are then left with $6$ solutions, namely
(we give here the full list of proto-triangles $x
\quad x\quad z\quad t\quad u\quad u$.)
 \begin{eqnarray*}
{<1,3,3>,\quad <1,3,3>,\quad <1,4,2>,\quad <1,2,4>,
\quad <2,2,3>,\quad <2,2,3>}\\
{<1,1,5>,\quad <1,1,5>,\quad <1,2,4>,\quad <1,4,2>,
\quad <1,3,3>,\quad <1,3,3>}\\
{<1,1,5>,\quad <1,1,5>,\quad <1,2,4>,\quad <1,4,2>,
\quad <2,2,3>,\quad <2,2,3>}\\
 {<1,1,5>,\quad <1,1,5>,\quad <1,4,2>,\quad <1,2,4>,
\quad <2,2,3>,\quad <2,2,3>}\\
{<1,3,3>,\quad <1,3,3>,\quad <1,2,4>,\quad <1,4,2>,
\quad <1,1,5>,\quad <1,1,5>}\\
{<1,3,3>,\quad <1,3,3>,\quad <1,2,4>,\quad <1,4,2>,
\quad <2,2,3>,\quad <2,2,3>}\\
\end{eqnarray*}
It is easy to see that the set of solution is invariant
 by the transformation $\pi
\leftrightarrow \tilde\pi$. We are finally left with the three solutions
\begin{eqnarray*}
 \alpha \quad \alpha \quad \pi \quad \tilde\pi \quad \beta \quad \beta \\
 \gamma \quad \gamma \quad \pi \quad \tilde\pi \quad \alpha \quad \alpha \\
\gamma \quad \gamma \quad \pi \quad \tilde\pi \quad \beta \quad \beta
\end{eqnarray*}
The other three solutions are obtained from the previous ones
 by taking their mirror images.
In this example of $A_6$ there is no further ambiguity in
the dissection process; in particular, the choice of one
of the two equal angles of isosceles triangles
determining a particular cutting is fixed by the
requirement that, depending upon the case, and as specified
by the above table of solutions, either
$\pi$ or $\tilde\pi$ should appear has an offspring. This is not a
generic feature valid for all values of $n$ (it is
even wrong when $n=4$ as we shall see below).\par

The solution $\alpha \alpha \pi \tilde\pi \beta \beta $ is illustrated in
Fig.5 (steps $t=1$, $t=2$, $t=3$, $t=9$, $t=10$).\par The solutions
$\gamma \gamma \pi \tilde\pi \alpha \alpha $ and $\gamma \gamma \pi
\tilde\pi \beta \beta $ give rise to analoguous pictures.\par Fig.16
shows in particular how $352$ triangles of type $\alpha$, $197$
triangles of type $\beta$ and $441$ triangles of type $\tilde \pi$
fit together to build a partition (line $t=11$ of the Brattelli
diagram of Fig.2) of their common ancestor, a ``big'' triangle that
is itself a union of one $\alpha$, one $\beta$ and one $\pi$ and
corresponds to the line $1$ of the same Brattelli diagram. All these
pictures were generated by a computer program written using
MATHEMATICA.\par Notice that no solution involves all five triangles
$\alpha$, $\beta$, $\gamma$, $\pi$ and $\tilde\pi$
simultaneously.\par

We discussed above the case associated with the Dynkin diagram of
$A_n$ with $n=6$. For other values of $n$, one proceeds exactly
in the same way. First, one establishes the list of possible
triangles (with angles that are integer multiple of
$\pi/(n +1)$) by looking at partitions of $n+1.$ Then, by solving the
equations for constraints, associated with the corresponding
Brattelli diagram, one chooses a solution and, accordingly, selects
$n$ types of triangles. Actually, when $n=2m$ is even, because the
$\Z_2$ symmetry of the Coxeter-Dynkin diagram, one needs only $m$
types of triangles along with their mirror image --this makes no
difference when the triangle is isosceles! Therefore $m$ types will
appear at even times $2 t$ and their mirror images will appear at odd
times $2 t+1$. Since all solutions to constraint equations appear
along with their mirror images, one can shorten the total list of
solutions for constraints (it is enough to give a reduced list).\par

Notice that discussion of the case $n=3$ is very simple. The corresponding
Coxeter-Dynkin graph and associated Brattelli diagrams are shown in
Fig.6. The associated dissections involve only isosceles triangles
with angles $\langle \pi/4, \pi/4, \pi/2 \rangle$ since $l=n+1=4 = 1
+ 1 + 2$. At time $t$, the ancestor triangle contains $2^t$ small
triangles (if $t$ is even) and $2^t + 2^t$ if $t$ is odd. The alert
reader recognizes the dimensions of the complexified Clifford
algebras. We shall return to that in the next section. Locally the
dissected ancestor triangle looks like part of a square tiling of the
plane (see Fig.7).\par

The case $A_4$ is
related with Penrose aperiodic tiling of the plane.
Corresponding Coxeter-Dynkin graph and associated
Brattelli diagrams are shown in Fig.8. The possible
triangles involved are obtained by writing
$l=n+1=5=3+1+1=1+2+2$. We set $\alpha = \langle 2,2,1
\rangle$ and $\beta =  \langle 1,1,3 \rangle$. Elementary
disintegrations are $\alpha \rightarrow \alpha + \beta$
and  $\beta \rightarrow \alpha + \beta$ The constraints
displayed in Fig.9 lead to the following equations
 \begin{eqnarray*}
x \rightarrow    x + y \\
y \rightarrow    x + y
\end{eqnarray*}
The solution is obviously $x=\alpha$ and $y=\beta$, or the converse.
The dissection of the ancestor triangle (of type $\alpha$ for
example) is however not entirely determined by the previous rule.
Indeed, there are two possible ways of cutting triangle
$\alpha$ (and two others for triangle $\beta$). More generally, once
the triangles are oriented, one could change the cutting rule at
each step. In other words, a solution to the set of
constraints specified  by a given Brattelli diagram does not always
specify the cutting rules (there was no ambiguity for the solutions
associated with the $A_6$ diagram but there are different
choices for $A_4$). Two such possible
choices are illustrated now: One can cut the first or the second
angle (both equal to $2 \pi/5$) of $\langle 2,2,1 \rangle$. These two
possibilities read (we label vertices by capital letters $\alpha =
ABC$, $\beta = DEF$, along with associated angles as a subscript, in
units of $\pi/ 5$, $M$ and $N$ are the new vertices appearing in the
dissection of $\alpha$ and $\beta$.)
 \begin{eqnarray*}
\langle A_1 B_2 C_2 \rangle &\rightarrow \langle C_1 M_1 B_1 \rangle +
\langle C_1 M_3 A_1 \rangle \\
\langle D_3 E_1 F_1 \rangle &\rightarrow \langle D_1 E_1 N_3 \rangle +
\langle D_2 N_2 F_1 \rangle
\end{eqnarray*}
or
 \begin{eqnarray*}
\langle A_1 B_2 C_2 \rangle &\rightarrow \langle A_1 B_1 M_3 \rangle +
\langle B_1 C_2 M_2 \rangle \\
\langle D_3 E_1 F_1 \rangle &\rightarrow \langle D_1 E_1 N_3 \rangle +
\langle D_2 N_2 F_1 \rangle
\end{eqnarray*}
Whatever dissection we choose as a starting point,
propagating this choice downward to all levels of the
infinite Brattelli tower, will lead at level $2t+1$ to two
kinds of triangles, $F_{2t+1}$ of type $\alpha$ and
$F_{2t+2}$ of type $\beta$, where $F_n$ are Fibonacci
numbers ($F_n=F_{n-1}+F_{n-2}$, $F_0=0, F_1=1$).
 Choosing an ancestor triangle of type $\alpha$, two
such infinite dissections are illustrated at level $t=13$ in Fig.10
and Fig.11 ($F_6=377$ triangles of type $\alpha$
 and $F_7=610$ triangles of type $\beta$). Locally, the fractal
triangular dissection of type
$A_4$ illustrated in Fig.10 looks like part of a Penrose
tiling of the plane (the one illustrated in Fig.11 does not).\par
In the case of Fig.10, calling $s_k=\sin{k \pi \over 5}$, $c_k=\cos{k
\pi \over 5}$, $\phi=4 \cos^2 {\pi \over 5} - 1 = 2 \cos {\pi\over
5}={1 + \sqrt 5 \over 2}=1.618\ldots$ (the golden number) and $\hat
\phi = -1/\phi = 1 - \phi$, we notice that, if we normalize the area
of the ancestor triangle (of type $\alpha$) to $1$, then, at level
$p$ (take it odd), all $F_p$ triangles of type $\alpha$ have the same
area, namely $(s_1/s_2)^{p+ 1}$, and all $F_{p+1}$ triangles of type
$\beta$ have also the same area, namely $(s_1/s_2)^{p-2} {c_2/ c_1}$.
We can check unitarity for all values of $t$ by writing $\Sigma \
{\rm Area}(\alpha$) + $\Sigma \ {\rm Area}(\beta) = 1$.  When
$t\rightarrow \infty$, we see that $\Sigma {\rm Area}(\alpha) =
-{\hat\phi / \sqrt 5} = {\phi - 1 / \sqrt 5}$ and $\Sigma {\rm
Area}(\beta) = 1 + {\hat\phi \over \sqrt 5}$ (so that, of course, the
total area is one).\par

The fact that all triangles of a given type
have the same area at a given step of the dissection is a
property of the case $n=4$  but
is not a generic feature of the constuction. In the
example, $n=6$, for example, we saw that such triangles are congruent
(similar) but not isometric in the Euclidean plane (see for instance
Fig.5). This property (the fact that one can make a partition of all
triangles appearing at a given step into a finite
number of types, all triangles belonging to a given type being
congruent) is of course true, by construction, for all values of
$n$. At this point, it may be useful to remember that
that $2\cos\pi/5$ is solution of a quadratic equation
($x^2-x-1=0$) whereas $2\cos\pi/7$ is solution of cubic equation
($x^3-x^2-2x+1=0$). \par

 Triangle dissections for bigger values of $n$
can be handled in a similar way. However, the combinatorics becomes
quite involved. For instance discussion of the case $n=8$ involves
$10$ different types of triangles (with angles that are multiples of
$\pi/9$) and the resolution by hand of the set of constraints is
quite painful. The following table gives the set of solutions
obtained thanks to a PROLOG-III computer program  (we list the
solutions $(a1=a2,a3,a4,a5,a6,a7=a8)$ by following the connected
vertices of the Coxeter-Dynkin diagram and  only give solutions
involving no repetitions of triangle types, i.e. $a1\neq a3,a1\neq
a5,a1\neq a7,a1\neq a4,a1\neq a6,a3\neq a5,a3\neq a7,a4\neq a6,a4\neq
a7,a6\neq a7$).
 \begin{eqnarray*}
 { <1,1,7>,
 <1,1,7>,  <1,6,2>,  <1,2,6>,  <2,4,3>,  <2,3,4>,  <2,2,5>,  <2,2,5>}\\
 { <1,1,7>,  <1,1,7>,  <1,6,2>,  <1,2,6>,  <1,3,5>,  <1,5,3>,  <1,4,4>,
 <1,4,4>}\\
  { <2,2,5>,  <2,2,5>,  <2,3,4>,  <2,4,3>,  <1,3,5>,
<1,5,3>,  <1,4,4>,  <1,4,4>}\\
 { <1,1,7>,  <1,1,7>,  <1,2,6>,  <1,6,2>,
<2,3,4>,
 <2,4,3>,  <2,2,5>,  <2,2,5>}\\
 { <1,1,7>,  <1,1,7>,  <1,2,6>,
<1,6,2>,  <1,5,3>,  <1,3,5>,  <1,4,4>,  <1,4,4>}\\
{ <2,2,5>,  <2,2,5>,<2,4,3>,  <2,3,4>,
<1,5,3>,  <1,3,5>,  <1,4,4>,  <1,4,4>}\\
\end{eqnarray*}
 Other cases of $A_n$ can be studied in the same way. One can in
particular look at the interesting cases $n=11$, $n=17$ and $n=29$
because of their relations with exceptional groups $E_6$, $E_7$ and
$E_8$ (they have same Coxeter numbers, namely $12, 18$ and $30$). \par
 \bigskip
 \sect{  Jones
algebras and triangular dissections}
Jones algebras appeared as an important tool in the study
of subfactors of Von Neumann algebras.
It will be enough, for us, to give the following
definition. $\tau$ being a real number, we call ${\cal
A}(\tau)$ the involutive algebra defined by an infinite
number of self adjoint generators $\{e_1, e_2, \ldots,
e_p, \ldots\}$ with relations
\begin{eqnarray*}
 e_p . e_p & = & e_p \\
e_p . e_{p\pm 1} . e_p & = & \tau e_p \\
e_p . e_q & = & e_q . e_p  \quad {\rm if} \ \vert q-p \geq 2 \vert
\end{eqnarray*}
Notice that the $e_p$'s are projectors. The
condition of being self-adjoint is actually quite strong;
it implies that ${1/ \tau}$ can not be arbitrary but must
belong to the following set \cite{Jones1}: $$
{1\over \tau} \in \{ 4 \ {\cos^2\pi/(n+1)}_{n\in \{2,3,4\ldots\}}\}
\cup \quad \lbrack {4}, \infty \lbrack
 $$ The quantity $Q \doteq 1/\tau$ is called the index.
 If one relax this condition of being
self-adjoint, one obtains algebras for all values of
$\tau$;  these algebras were introduced by
Temperley and Lieb \cite{TEMP}; we shall
call them Jones algebras since the $*$-condition is,
for us, essential. Our triangular dissections are
actually related to Jones algebras with index $1/\tau$ in
the discrete range
${\tau} \in \{ [4 \ {\cos^2\pi/(n+1)]^{-1}\}}_{n\in \{2,3,4\ldots\}}$. Readers
knowing Clifford algebras will have recognized that, in the case $\tau =
1/2$, one can set $e_p = 1/2 (1+\gamma_p \gamma_{p+1})$
where the $\gamma$'s generate the usual (infinite
dimensional) Clifford algebra: $\gamma_p^2=1$, $\gamma_p
\gamma_q + \gamma_q \gamma_p = 0$ if $p \neq q$. The
algebra generated by the $e_p$ then coincides, when $\tau=1/2$
with the even part of the infinite dimensional
Clifford algebra. Looking at the infinite Brattelli
diagram of $A_3$ displayed in Fig.6 (one recognizes the
dimensions of finite dimensional Clifford algebras)
suggests then an alternative definition of Jones algebras
(for general $\tau$) which is totally explicit and
actually better suited for our purposes.\par
 We therefore consider the (infinite)
Bratelli diagram as describing an infinite tower of matrix
algebras, with embeddings specified by the edges on the
diagram (this is actually why such diagrams were
introduced in the first place \cite{Bra}). For example, in the
case of $A_4$, i.e.
$\tau = {4 \cos^2 \pi/5}^{-1} = \phi^2$ ($\phi$ is the golden number), the
lines $t=3$, $t=4$ read as Fig.12. The first line describes a direct
sum of two matrix algebras $M_2(\C) \oplus M_3(\C)$ embedded in the
matrix algebra $M_5(\C) \oplus M_3(\C)$ as Fig.13. The embedding was
here quite easy to describe since $A_n$ diagrams involve only single
lines but this can be generalized. For a given value of the discrete
time flow $t$ (i.e. at the line $t$), we have an algebra ${\cal
A}_t(\tau)$ that is a direct sum of finite dimensional
algebras with dimensions specified by the integers
appearing at level $t$. The Jones algebra is actually
defined as the inductive limit of this sequence of
algebras. We refer to \cite{Jones1} \cite{Jones2} \cite{JON-HAR}
\cite{Connes-Jones} for more
information, but we can summarize the situation as follows. An essential
ingredient of the construction is the existence of a
trace $Tr$. The first remark is that, on a simple
finite dimensional matrix algebra, the trace is unique,
up to scale. This is not so on a direct sum of simple
finite dimensional matrix algebras since any linear
combination of component traces is a trace (in the sense
that $tr(a.b)=tr(b.a)$). In the case of an infinite
family such as the one defined by the Brattelli diagram
of $A_4$, it is possible to define a trace at each line
$t$ of the diagram, in such a way that it is compatible
with the embedding defined by the diagram. This property
allows one to define a ${\sl unique}$
trace $Tr$ on all the algebras ${\cal A}_t(\tau)$. For
example, in the case $n=4$, i.e. $\tau=\phi^2$, the
trace $Tr$ can be explicitly defined at level $2t+1$,
 where ${\cal A}_t(\tau) = M_{F_{2t+1}}(\C)
\oplus M_{F_{2t+2}}(\C)$ by $$ {\rm Tr} = \tau^{t+1} {\rm tr}_\alpha +
\tau^{t} (1-\tau) {\rm tr}_\beta$$ and at level $2t$, where ${\cal
A}_t(\tau) = M_{F_{2t+1}}(\C) \oplus M_{F_{2t}}(\C)$ by
$$ {\rm Tr} = \tau^{t}  {\rm tr}_\alpha + \tau^{t-1} (1-\tau)
{\rm tr}_\beta$$ where $tr_\alpha$ and $tr_\beta$ denote the ``naive''
traces on the corresponding simple factors. The reader can verify
that such a definition is indeed compatible with the specified
embeddings. It can then be checked that $$ Tr(x e_{n+1}) = \tau
Tr(x)$$ if $x$ belongs to the algebra generated by $1,e_1,e_2\ldots
e_n$. In particular $Tr(e_n)=\tau$. This trace $Tr$ is also used to
give a topology of $\C^*$ algebra to the inductive limit. This is the
Jones algebra ${\cal A}(\tau)$. Its weak closure is a Von
Neumann factor of type ${\rm II}_1$ (there is a trace!) \par

The trace of projectors obviously belongs to
the set $\Z \oplus \Z$ (in the case $n=4$) or to the set $\Z
\oplus \ldots \oplus \Z$ ($2 m$ times) in the case of
$A_{2m}$. This discussion leads to the study of the
$K$-theory of algebras ${\cal A}(\tau)$ and we refer to
\cite{Connes1} \cite{Connes2} for a discussion of the case $\tau =
\phi^2$ where it is shown that $K_0 ({\cal
A}(\tau)) = \Z \oplus \tau \Z$. Notice that the first
two lines of diagram (Fig.12) describing the inclusions and
determining the infinite Brattelli diagram are fully
specified by the positive matrix $$
\pmatrix{ 1 &1 \cr 1 &0 \cr}$$
and that the largest eigenvalue of this matrix is
precisely equal to $\tau$. For more information about
Brattelli diagrams, graphs, incidence matrices and
Perron-Frobenius vectors, we refer to \cite{JON-HAR}.\par

We shall return a little bit later to Jones algebras but
notice already that they are naturally represented on
the fractal sets described in the previous section.
By this, we mean the following: Take for example $n=6$ (i.e
$\tau = 1/ (4\cos^2{\pi\over 7})$) and $t=4$; we can
decorate all $ 6 + 7 + 3 = 16$ triangles appearing at that
level with particular complex numbers (or colors) and put
this information in a column vector with $16$ components;
the matrix algebra ${\cal A}_{t=4}(\tau)=M_6(\C)\oplus
M_7(\C) \oplus M_3(\C)$ acts on the vector space $\C^{16}=\C^6
\oplus \C^7 \oplus \C^3$ spanned by such vectors. In other words,
the set of all triangles appearing at level $t$ provides a basis for a
representation of the algebra ${\cal A}_{t}$. The different sets of
triangles associated with a given type (there are $3$ of them in
this example) correspond to the irreducible representations of the
semi-simple algebra ${\cal A}_{t}(\tau)$.\par

\sect{Infinite sequences, non-commutative geometry
and aperiodic tilings}\vs{10}
\par
We start with the discussion of $A_4$. Notice first that a
tiling by Penrose kites and darts gives rise to a
tiling using the two kinds of triangles discussed in the
first section if we simply cut the kite and the dart into
two equal triangular pieces. \par
It is known (for example \cite{Gru}) that there exists an
infinite number (cardinality of the continuum) of
inequivalent (aperiodic) Penrose tilings of the plane. It
is also known (see the nice discussion in \cite{Gardner}) that
each finite part of a given tiling can be found (an
infinite number of times) in any other tiling. Each Penrose
tiling can be encoded by an infinite sequence of $0$ and
$1$ obeying the following rule: $1$ is isolated. Such a
sequence gives actually a construction procedure for a
Penrose tiling (sometimes called a ``Penrose universe'').
 Two sequences that differ only by a finite number of
terms define the same tiling (only the chosen ``origin''
is different).  Two such sequences
are declared equivalent when they differ by a finite
number of terms (i.e. they agree after some rank $t$).
The space of equivalence classes is then a Cantor set
(measure $0$ and cardinality of $\aleph_1 = 2^{\aleph_0}$). The above
are standard results. We shall reinterpret them below.\par
Non-commutative geometry, as described in \cite{Connes2} gives us a
new way of encoding equivalence classes (or more generally
groupoids); this description uses (non-commutative) algebras rather
than (quotient) sets. In the case of finite dimensions, rather than
collapsing an equivalence class (with $p$ elements) to a single point,
one describes it by a full matrix algebra $M_p(\C)$. In
the present case of Penrose tilings, one \cite{Connes2}
declares that two finite sequences of $0$ and $1$ (satisfying
the rule ``$1$ is isolated'') are equivalent if they end by
the same digit. It is easy to check that there are $F_t$
such sequence of length $t$ ending  by $0$ and
$F_{t-1}$ such sequence of length $t$ ending by $1$. For
each value of $t$ we have therefore two equivalence
classes; according to the philosophy of non-commutative
geometry, they are encoded by the algebra $M_{F_t} \oplus
M_{F_{t-1}}$ i.e. by ${\cal A}_t(\tau)$, with $\tau =
\phi^2$. There is a projection from the space of
sequences of length $t$ to the space of sequences of
length $t-1$; it is obtained by erasing the last digit.
One can see that, correspondingly, there is an inclusion
from the algebra  ${\cal A}_{t-1}(\tau)$ into the algebra
${\cal A}_t(\tau)$. This inclusion coincides with the one
described by the Brattelli diagram. One can (and must) of
course consider infinite sequences: Penrose tilings are
described by equivalence classes of infinite sequences.
 Corresponding to the projective limit of
sequences we have an inductive limit of algebras. In other words,
non-commutative geometry encodes the space of inequivalent
Penrose tilings by the Jones algebra ${\cal A}({\phi^2})$.
 This is, in essence the message of
chapter II-3 of the book \cite{Connes2}, see also
\cite{Connes1}.\par An infinite sequence of the previous
kind also describes a never ending path on the Brattelli
diagram, i.e. an infinite dive within the fractal set
obtained by  infinite dissection in the first part of this
article. The correspondence between sequences of triangles
and sequences of $0$ and $1$ will be discussed shortly.
Using triangular dissection to generate
Penrose tilings is a known technique (it was advocated for
example in the book \cite{Mathematica}). What is important to notice
here is that the fractal triangle of type $A_4$ constructed in
section 2 contains, in a sense, {\sl all} Penrose tilings; it is, in
a sense, analogous to the Mandelbrot set that encodes
all possible Julia sets. Indeed, an infinite dive in this
fractal triangle describes a particular Penrose tiling. A
finite dive (at depth $t$) corresponds to the choice of a
specific triangle appearing at a finite value of $t$. If we
inflate this triangle by the proper scale (some power of the
golden number), it can be super-imposed onto the first
``ancestor-triangle'' and we obtain in this way an
aperiodic tiling of part of the plane. Equivalently, we
can ``cut a window'' around the chosen small triangle and
inflate it. The fact that following a path along the
Brattelli diagram (or choosing triangles within triangles)
ultimately defines a tiling of the plane is ensured by the
hierarchical nature of the construction and in particular
by the compatible embeddings of all algebras
${\cal A}_t(\tau)$, i.e. by the definition and existence of
${\cal A}(\tau)$ as an inductive limit.\par
Now, there is nothing special about $n=4$. Other ``fractal
triangles'' corresponding to Coxeter-Dynkin graphs of type
$A_n$ have been constructed in the first section and they
play the same role as above. As in the case of $A_4$ it
may be useful to define directly new kinds of infinite
sequences with appropriate constraints along with an
equivalence relation, in such a way that the Jones algebra
${\cal A}(\tau)$ appears {\sl a priori} as the
non-commutative object encoding the space of equivalence
classes. Let us give the construction rules for such
sequences in a few cases. The simplest of all is certainly the case
$A_3$ that is encoded by the Clifford algebras ($\tau = 1/2$).
The associated finite sequences of length $t$ are just sequences of
$0$ and $1$ with no particular constraint. Two finite sequences are
equivalent if they end with the same digit and there are clearly
$2^{t-1}$ such equivalent sequences at level $t$. \par
For a general bicolored Dynkin-Coxeter graph, the following technique
obviously works (but as we shall see it can be improved in the case
of $A_{2m}$): We label red (say) vertices by integers $1, 2, 3 \ldots$
and green (say) vertices by letters $\alpha, \beta, \gamma \ldots$. We
build sequences starting by $1, 2, 3 \ldots$ containing both integers and
greek letters with the following constraint: a symbol $x$ can be
followed by the symbol $y$ if $x$ and $y$ are linked on the
Dynkin-Coxeter graph. The numbers of equivalent finite sequences
(i.e. sequences ending by the same symbol) of given length are given
by the integers appearing on the associated Brattelli diagram. For
example, in the case of $A_4$ and for $t=4$, we obtain the following
words $1 \alpha 1 \alpha$, $1 \alpha 2 \alpha$, $2 \alpha 1 \alpha$,
$2 \alpha 2 \alpha$, $2 \beta 2 \alpha$ and $1 \alpha 2 \beta$, $2
\alpha 2 \beta$, $2 \beta 2 \beta$, i.e. $3$ sequences ending with
$\beta$ and $5$ sequences ending with $\alpha$ (Fibonacci numbers, of
course). But we already know that, in the case of $A_4$ it is enough
to consider sequences containing only two symbols ($0$ and $1$) that
obey the rule: $1$ is isolated. More generally and thanks to $\Z_2$
symmetry, the rules for $A_n$ can be simplified when $n$ is even
($n=2m$). In that case, we fold the Dynkin-Coxeter
 graph ``over itself'' and obtain
a string of $m=n/2$ vertices, the first one being related to itself by
a closed loop (see Fig.14 for the case of $A_6$). We then label the
vertices from $0$ to $m-1$ and decide to build sequences that start by
integers taken from the set $0,1\ldots,m-1$ and that obey the following
constraint: the symbol $x$ can be followed by the symbol $y$ if they
are linked (neighbors) in the list $0 \leftrightarrow 0 \leftrightarrow 1
\leftrightarrow 2 \leftrightarrow \ldots \leftrightarrow m-1
$. For instance the constraints for $A_6$ read
$0\leftrightarrow 0 \leftrightarrow 1 \leftrightarrow 2$. It is a
simple combinatoric exercise to check that the number of finite
sequences of length $t$ obeying the previous rules and ending by the
same given digit is again given by the integers appearing at depth
$t$ on the associated Brattelli diagram. Such sequences of length $t$
encode paths of the Brattelli diagram starting at the top level $t=1$
and ending at depth $t$. The correspondence goes as follows: We again
draw the Brattelli diagram but this time, we label vertices from
left to right with integers $0,1,2\ldots m-1$ when $t$ is odd but
with integers $m-1, m-2,\ldots, 1,0$, again from left to right, when
$t$ is even (see Fig.15 in the case of $A_6$). A sequence of the
previous kind clearly labels a path from top to bottom by following
the corresponding integers (the path $0, 2, 0, 2, 1, 1$ is highlighted
on Fig.15). This sequence, in turn, specifies a particular triangle
of the dissection process. We have seen that the Jones algebra
associated with some value of $\tau$ acts on a vector space for which we can
choose a basis labeled by the space $X$ of triangles of the
dissection process. We can of course replace these triangles by paths or by
sequences. What is nice with triangles is that one can ``see'' them!
Before ending this sub-section devoted to sequences encoding paths
and their relations with triangular dissections, let us remember that
Fibonacci numbers (i.e. case of $A_4$) define the Fibonacci number
system \cite{Knuth}: We can represent any nonnegative integer $n$ as a
sequence of $0$'s and $1$'s writing $$
n=(b_m b_{m-1}\ldots b_2) \Leftrightarrow n=\Sigma_{k=2}^m b_k F_k
$$ This system is like binary notation, except that there are never
two adjacent 1's. For instance $7=5+2=F_4+F_2=(1010)$. Incidentally
this property allows one to give a particular total ordering to the space of
finite paths of the Brattelli tower of $A_4$, or equivalently, to
the space of triangles appearing after a finite number of
dissections.The other sequences of type $A_k$ (dimension of finite
dimensional Jones algebras) for all $k$ appear like generalizations
of the Fibonacci number system.
 Notice that it is possible to give an algebra structure to the set
of discrete  strings (paths) living on the kind of diagrams that we
discuss in the present paper \cite{Ocneanu}; this
algebra structure could be discussed in terms of our
triangular dissections but we did not this subject here.

\par

\sect{ Remarks about the projection technique}

We have seen in the first part of this article how to generalize the
dissection techniques known to generate Penrose aperiodic tilings of
the plane \cite{Mathematica} and indexed by the golden number. This
generalization gives triangular dissections indexed by other
algebraic integers belonging to the family $4 \cos^2{\pi\over n+1}$.
Specialists of quasicrystals with pentagonal symmetry know that there
is another quite powerful technique to generate Penrose tilings: The
so-called projection technique \cite{Krammer} (or dimensional
reduction technique). This technique has already been generalized to
other cases \cite{Duneau}, \cite{Pleasant},\cite{Pleasant2}. One can
start with  $\R^p$, ($p=5$ in the case of Penrose tilings) decompose
$\R^p$ into eigenspaces of the matrix $\rho$ representing the
generator of the group $\Z_p$. It is actually convenient to use a
self adjoint matrix like $M = \rho + \rho^\dagger$, so that we obtain
a decomposition of $\R^p$ into eigenspaces corresponding to real
eigenvalues. One then chooses a real decomposition of $\R^p$ into
eigenspaces $\R^p= D \times T \times W1 \times W2 \times \ldots $
where $D$ is associated with eigenvalue $2$, and such that $M$ acts
on $T$ by dilatation (eigenvalue $>1$). From the definition of $M$
(and the fact that $\rho^p=1$), it is clear that spaces $T$, $W_1$,
$W_2 \ldots$ are all $2$-dimensional.
 We choose the regular lattice $\Z^p$ (actually some finite part of it
contained in a hypercube) and project $\Z^p$ on the space $T \times W1
\times W2 \times \ldots $ parallely to $D$. In this way we obtain a
regular lattice related to the root lattice of $A_{p-1}$. We then
choose several $2p$-gonal windows in the window spaces $W_1,W_2
\ldots$ and keep only those points that can be projected (parallely
to the sum of the other eigenspaces) simultaneously within these
windows (i.e. the points belonging to the common shadow of all
the windows). The last step is to project the remaining points on the
two-dimensional target space $T$. The necessity of choosing windows
before projecting on $T$ comes from the fact that the image would
otherwise be dense. The obtained discrete point set has many
``good'' properties, in particular, it is crystal-like and
characterized by an inflation multiplier $\tau$. Many pictures of
quasicrystals obtained by this technique have been published in the
case of $\Z_5$. We have followed this technique in the case of $\Z_7$ and
obtained Fig.17. Here we set $M=-(g^2+g^5)$, with $g^7=1$, approx.
eigenvalues  of $M$ are $   {-2., 1.80194, 1.80194, -1.24698,
-1.24698, 0.445042, 0.445042}$, the two-dimensional target $T$ is
associated with $2\cos\pi/7 =1.80194\ldots$ and the two other window
spaces $W_1$ and $W_2$ with the two other eigenvalues; we start with
$4^7=16384$ lattice points in $R^7$, the intersection of the
``shadows'' in $R^6$ of two 14-gonal windows centered in $W_1$ and in
$W_2$ contains $2119$ points that are projected on $T$ i.e. in Fig.
17 (only lattice points have been drawn). We have also considered the
choice $M=-(g^2+g^3+g^4+g^5)$ and obtained very similar pictures.
There is no direct correspondence between the number and locations
of points (vertices of triangles) obtained after a dissection
 of $t$ steps and the points
obtained via the dimensional reduction technique. Indeed, the ``order of
arrival'' of points obtained thanks to the projection technique is a
function of the arbitrarily chosen size of windows and of the size of
the chosen hypercube of $\R^5$.  The dissection technique discussed
previously looks more fundamental, in the sense that one can follow
step by step the inductive algebraic process and the role of Jones
algebras. It is however quite interesting to superimpose Fig.17 (projection
technique) with Fig.18 (triangular dissection technique) both
obtained in the study of the $A_6$ case.\par
The use of $K$-theory in the coding of aperiodic lattices has been advocated
several times \cite{Connes1},\cite{Kellendonk}; such structures can be
obtained thanks to a projection technique but the direct link
with the inductive algebraic process -- described by Brattelli diagrams-- is
then lost.\par
The fractal aperiodic structures discussed in the present paper cannot be
called ``tilings''  (since these triangular dissections usually involve
triangles that are congruent but not isometric). The fact that they can be
related to  ADE Coxeter-Dynkin diagrams may come as a surprise and is maybe
another of the many incarnations of Gabriel's theorem
\cite{Gabriel}\cite{Gelfand}. \par
The last Figure (Fig.18) is obtained by rotating a triangular dissection of
order seven and could, maybe, inspire children willing to play a fractal
version of the game of hopscotch (``marelle'').

\indent

{\Large{\bf{Acknowledgments}}}

\indent

I want to thank the members of the Department of Mathematics of the
University of Macquarie (Sydney) where this work started, in
particular Prof. J. Corbett and M. Adelman for their
invitation. I also thank my colleagues from the Centre de Physique
Theorique (in particular G. Esposito Farese), in Marseille, for their
interest and useful comments.

\vs{4}

\newpage

{\bf Figure Captions}
\begin{enumerate}
\bigskip
\item  The bicolored Coxeter-Dynkin graph of $A_6$
\item  The Brattelli diagram associated with the graph of $A_6$ (Jones
algebra of index $4 \cos^2 \pi/7$)
\item  A regular heptahedron displaying triangles of types $\alpha, \beta,
\gamma, \pi, \tilde\pi$
\item  The constraint equations for triangular dissections of type $A_6$
\item  Dissection of a triangle $\langle 4, 2, 1\rangle \ \pi/7$
according to the embeddings specified by the Brattelli diagram
associated with a bicolored Coxeter-Dynkin graph of $A_6$.
\item  The Brattelli diagram associated with the Coxeter-Dynkin graph of
$A_3$  (Jones algebra of index $4\cos^2\pi/4=2$ i.e. Clifford
algebra)
\item  Case of $A_3$. Dissection of a triangle $\langle 1, 1, 2\rangle
\ \pi/4$ into a union of $64$ triangles. Step $t=6$.
\item  The Brattelli diagram associated with the Coxeter-Dynkin graph of
$A_4$ (Jones algebra of index $4\cos^2\pi/5=1/\phi^2$)
\item  The constraint equations for triangular dissections of type $A_4$
\item  Case of $A_4$. Penrose-like structure obtained by dissection
of a triangle $\langle 2, 2, 1\rangle\ \pi/5$. There are $610 +
377$ triangles (Fibonacci numbers) belonging to two different
families. Step $t=14$.
\item  Same as Fig.10 but with a different choice for the elementary
matching conditions.
\item  Part of the Brattelli diagram associated with a bicolored
Dynkin-Coxeter graph of type $A_4$.
\item  Embedding of algebras described by the diagram of Fig.12.
\item  Diagram obtained by folding the graph of $A_6$ over itself. It
describes sequences obeying the rule
$0\leftrightarrow 0 \leftrightarrow 1 \leftrightarrow 2$
\item  A path on the Brattelli diagram of $A_6$ caracterized by the
sequence $0-2-0-2-1-1$.
\item   Case of $A_6$. Partition of a triangle $\langle 4, 2, 1\rangle \
\pi/7$ as a union of $197+441+352$ triangles belonging to three
congruent families. Step $t=11$.
\item  A particular $2$-dimensional projection of the root lattice of
$A_6$. The projector is obtained from the decomposition of $R^7$
into eigenspaces of $\rho + \rho^\dagger$ with $\rho^7 = 1$. The
original points belong to the shadow of a four dimensional window
which is the product of two $14$-gones.
\item  Picture obtained after symmetry and $7$-fold rotation of the
$A_6$-fractal dissection of a triangle $\langle 4, 2, 1\rangle \ \pi/7$.
\end{enumerate}
\end{document}